\documentclass[]{interact}

\usepackage{epstopdf}
\usepackage[caption=false]{subfig}
\usepackage{pdflscape}

\theoremstyle{plain}

\theoremstyle{definition}

\theoremstyle{remark}

\begin{document}


\title{Lie symmetries and similarity solutions for a family of 1+1
fifth-order partial differential equations}

\author{
\name{Andronikos Paliathanasis\textsuperscript{a}\thanks{ Email: anpaliat@phys.uoa.gr}}
\affil{\textsuperscript{a}Institute of Systems Science, Durban University of Technology, PO Box 1334, Durban 4000, Republic of South Africa}
\affil{Instituto de Ciencias F\'{\i}sicas y Matem\'{a}ticas, Universidad Austral de Chile, Valdivia, Chile}
}

\maketitle

\begin{abstract}
We apply the theory of infinitesimal transformations for the study of a
family of 1+1 fifth-order partial differential equations which have been
proposed before for the description of multiple kink solutions. In this
analysis we perform a complete classification of the Lie symmetries and of
the one-dimensional optimal system. The results are applied for the
derivation of similarity solutions and in particular we find travel-wave and
scaling solutions. We show that the kink-solution of these equations can be
recovered by the use of the Lie symmetries, while new solutions are also
derived.
\end{abstract}


\section{Introduction}

In a series of books, Sophus Lie established the theory of infinitesimal
transformation for the analysis of differential equations \cite%
{lie1,lie2,lie3}. In particular Lie's work was to take the infinitesimal
representations of the finite transformations of continuous groups, thereby
moving from the group to a local algebraic representation, and to study the
invariance properties under them. This resulted in linearization of all
equations and/or functions under consideration.

The basic purpose for the determination of the invariant transformations
which leave invariant a given differential equation, known as Lie
symmetries, it is to facilitate the solution of the given differential
equations. The existence of a sufficient number of Lie symmetries of the
right type, for a differential equation, enables one to solve the
differential equation by means of repeated reduction of order and a reverse
series of quadratures or by means of the determination of a sufficient
number of first integrals. There are differences in the reduction process
between ordinary and partial differential equations. For ordinary
differential equations the application of a Lie point symmetries leads to
the reduction of the order for the differential equation by one. On the
other hand for partial differential equations the application of a Lie point
symmetry leads to a new differential equation where the number of
independent and dependent variables, however, the order of the reduced
equation is the same with the original equation.

Lie symmetries have been applied in various problems of mathematical physics
and applied mathematics. The famous Noether theorem, is the application of
Lie's theory in the Action Invariant of the Calculus of Variations \cite%
{Noether18}. The main novelty of Noether's theorem is the direct relation of
the infinitesimal transformations which leave invariant the Action Integral
with conservation law for the equation of motions. Consequently, Lie's
theory is essential for various physical theories, from classical mechanics,
quantum mechanics, fluid theory and many other \cite%
{m1,m2,m3,m4,m5,m6,m7,m8,m9,m10}. Additionally, because the analysis of the
Lie symmetries is a systematic way for the determination of exact and
analytic solutions of nonlinear differential equations, it plays an
important role in all theories of applied mathematics, we refer the reader
in \cite{r1,r2,r3,r4,r5,r6,r7,r8,r9,r10,r11,r12,r13,r15,r16,r17} and
references therein.

In this work we are interest on the application of Lie symmetries for the
two 1+1 fifth-order partial differential equations \cite{kink1}

\begin{equation}
\Phi _{A}:u_{ttt}-u_{txxxx}-\alpha \left( u_{x}u_{t}\right) _{xx}-\beta
\left( u_{x}u_{xt}\right) _{x}=0,  \label{p.01}
\end{equation}%
\begin{equation}
\Phi _{B}:u_{ttt}-u_{txxxx}-u_{txx}-\alpha \left( u_{x}u_{t}\right)
_{xx}-\beta \left( u_{x}u_{xt}\right) _{x}=0,  \label{p.02}
\end{equation}%
which has been proposed before to describe (multiple-) kink solutions \cite%
{kink2,kink3}. Lie symmetries has been applied before in various systems or
the determination of solitons. The 2+1 sine-Gordon equations equations has
been studied by using the Lie symmetry method in \cite{kink4}. Solitons for
the (2+1)-dimensional Bogoyavlenskii equations were found as similarity
solitons in \cite{kink5}. For other applications we refer the reader in \cite%
{kink6,kink7,kink8,kink9,kink10} and references therein. The plan of the
paper is as follows.

In Section \ref{sec2}, we briefly discuss the mathematical tools of our
analysis, that is, we give the basic definitions of Lie's theory and we
discuss the importance for the determination of one-dimensional optimal
system. In addition, we present the basic elements of the singularity
analysis, because it is applied for the determination of analytic solutions
of the reduced equations. In Section \ref{sec3}, we present the admitted Lie
symmetries for the two equations of our study. We determine the construction
constants of the admitted Lie symmetries, as we present the elements of the
adjoint representation. We found some particular cases for the free
parameters $\alpha $ and~$\beta $, where the equations are invariant under a
different Lie algebra The application of the Lie invariants in presented in
Section \ref{sec4} where we determine travel-wave and scaling similarity
solutions. The behaviour of these solutions is discussed. Finally, in
Section \ref{sec6} we summarize our results and we draw our conclusions
while we discuss possible extensions of our analysis.

\section{Preliminaries}

\label{sec2}

In this section we briefly discuss the main mathematical tools that we apply
in this work. Specifically, we discuss the theory of infinitesimal
transformations known as Lie's theory, and the Painlev\'{e} analysis, also
knows as singularity analysis.

\subsection{Lie symmetries}

Let us consider the infinitesimal one-parameter point transformation, \
\begin{align}
\bar{x}^{k}& =x^{k}+\epsilon \xi ^{i}\left( x^{k},u\right) , \\
\bar{\eta}& =\eta +\epsilon \eta \left( x^{k},u\right) .
\end{align}%
with generator $X=\xi ^{i}\left( x^{k},u\right) \partial _{i}+\eta \left(
x^{k},u\right) \partial _{u}.$

The vector field $X$ it is called a Lie\ (point) symmetry of the
differential equation $\mathbf{H}\left(
x^{k},u,u_{i},u_{ij},...,u_{i_{1}i_{2}...i_{n}}\right) $, if there exists a
function $\lambda $ such that the following condition to be true \cite%
{ibra,Bluman,Stephani}%
\begin{equation}
X^{\left[ n\right] }\mathbf{H}-\lambda \mathbf{H=0}  \label{go.11}
\end{equation}%
where $X^{\left[ n\right] }$ is called the n-th prolongation/extension of $%
X~ $in the jet-space defined as%
\begin{equation}
X^{\left[ n\right] }=X+\left( D_{i}\eta -u_{,k}D_{i}\xi ^{k}\right) \partial
_{u_{i}}+\left( D_{i}\eta _{j}^{\left[ i\right] }-u_{jk}D_{i}\xi ^{k}\right)
\partial _{u_{ij}}+...+\left( D_{i}\eta _{i_{1}i_{2}...i_{n-1}}^{\left[ i%
\right] }-u_{i_{1}i_{2}...k}D_{i_{n}}\xi ^{k}\right) \partial
_{u_{i_{1}i_{2}...i_{n}}}.  \label{go.13}
\end{equation}

For any symmetry vector field $X$, we can define the canonical coordinates
such that $X=\partial _{x^{n}}$. It follows that the differential equation
admits the Lie symmetry~$X=\partial _{x^{n}}$ when~$\mathbf{H=H}\left(
x^{\mu },u,u_{i},u_{ij},...,u_{i_{1}i_{2}...i_{n}}\right) ~$,$~\mu \neq n.$

Lie symmetries are determined to find similarity transformations with the
use of the so-called Lie invariants. The algorithm which is applied for the
determination of the Lie invariants is to solve the following Lorentz system
\begin{equation}
\frac{dx^{i}}{\xi ^{i}}=\frac{du}{u}=\frac{du_{i}}{u_{\left[ i\right] }}=...=%
\frac{du_{ij..i_{n}}}{u_{\left[ ij...i_{n}\right] }}.
\end{equation}

Similarity solution for a differential equation is called the solution which
is determined by the use of the Lie invariants.

\subsection{One-dimensional optimal system}

Assume the $n$-dimensional Lie algebra $G_{n}$ with elements $\left\{
X_{1},~X_{2},~...~X_{n}\right\} ~$and structure constants $C_{jk}^{i}$. We
define the two symmetry vectors
\begin{equation}
Z=\sum_{i=1}^{n}a_{i}X_{i}~,~W=\sum_{i=1}^{n}b_{i}X_{i}~,~\text{\ }%
a_{i},~b_{i}\text{ are constants.}  \label{sw.04}
\end{equation}%
Vectors $Z$ and $W$ are equivalent if and only if \cite{olver}
\begin{equation}
\mathbf{W}=\sum_{j=i}^{n}Ad\left( \exp \left( \epsilon _{i}X_{i}\right)
\right) \mathbf{Z}  \label{sw.05}
\end{equation}%
or
\begin{equation}
W=cZ~,~c=const\text{ that is }b_{i}=ca_{i}\text{.}  \label{sw.06}
\end{equation}

The new operator%
\begin{equation}
Ad\left( \exp \left( \epsilon X_{i}\right) \right) X_{j}=X_{j}-\epsilon
\left[ X_{i},X_{j}\right] +\frac{1}{2}\epsilon ^{2}\left[ X_{i},\left[
X_{i},X_{j}\right] \right] +...  \label{sw.07}
\end{equation}%
is called the adjoint representation.$~$

For the Lie algebra $G_{n}$ and from the corresponding adjoint
representation we can calculate the invariants, which are the constants $%
a_{i}$ in the generic symmetry vector $Z$, that can not be eliminated. The
relative invariants are given by the solution of the following system of
first-order partial differential equations $\Delta ^{i}\left( \Phi \left(
a^{\rho }\right) \right) \equiv 0,$where the differential operators $\Delta
^{i}$ are defined as $\Delta ^{i}=C_{jk}^{i}a^{j}\frac{\partial }{\partial
a^{k}}.$

The determination of all the one-dimensional subalgebras of $G_{n}$ which
are not related through the adjoint representation it is necessary in order
to perform a complete classification of all the possible similarity
transformations, i.e. similarity solutions, for a given differential
equation. This classification is known as the one-dimensional optimal system.

\subsection{Singularity analysis}

The development of the Painlev\'{e} Test for the determination of
integrability of a given equation or system of equations and its
systematization been succinctly summarized by Ablowitz, Ramani and Segur in
the so-called ARS algorithm \cite{Abl1,Abl2,Abl3}. \ The three basic steps
of the ARS algorithm are (a) determine the leading-order term which
describes the behaviour of the solution near the singularity, (b) find the
position of the resonances \ which shows the existence and the position of
the integration constants and (c) write a Laurent expansion with
leading-order term determined in the first step in order to perform the
consistency test and the solution, for a review the ARS algorithm we refer
the reader in \cite{sin1}, while a recent discussion between the symmetry
analysis and the singularity analysis can be found in \cite{sin2}.

Let us demonstrate the application of the ARS algorithm by considering the\
Painlev\'{e}-Ince Equation \cite{pp1}%
\begin{equation}
y^{\prime \prime }+3yy^{\prime }+y^{3}=0.
\end{equation}%
For the first step of the ARS algorithm we $y\left( x\right) =a_{0}\left(
x-x_{0}\right) ^{p}$ in the equation, from where the Painlev\'{e}-Ince
equation becomes
\begin{equation}
a_{0}p\left( p-1\right) \left( x-x_{0}\right) ^{p-2}+3p\left( a_{0}\right)
^{2}\left( x-x_{0}\right) ^{2p-1}+a_{0}\left( x-x_{0}\right) ^{3p}=0.
\end{equation}

From the latter algebraic equation, balance occurs if $p=-1$ which gives
that $a_{0}=1$ or $a_{0}=2$. Hence there are two two possibilities of
leading-order behaviour.

The resonances are determined by substitute $y\left( x\right) =\left(
x-x_{0}\right) ^{-1}+m\left( x-x_{0}\right) ^{-1+s}$ in the original
equation and linearize around $m^{2}\rightarrow 0$. We end with the\
polynomial equation $\ \left( s-1\right) \left( s+1\right) =0~$ the solution
of which provides $s=-1$ and $s=1$. The value $s=-1$ is to be expected as it
is associated with the movable singularity. Consequently, the algebraic
solution of the Painlev\'{e}-Ince equation is given in terms of the Right
Painlev\'{e} series
\begin{equation}
y\left( x\right) =\left( x-x_{0}\right) ^{-1}+\sum\limits_{I=1}^{\infty
}a_{I}\left( x-x_{0}\right) ^{-1+I}.
\end{equation}%
For the coefficient $a_{0}=2$ of the leading-order term we find the
resonances $s=-1$ and $s=-2$ which means that the solution is given by the
Left Painlev\'{e} Series

\section{Lie symmetries and one-dimensional optimal system}

\label{sec3}

In the following we apply the Lie algorithm in order to determine the
generators of the infinitesimal transformations which leave invariant the
differential equations (\ref{p.01}) and (\ref{p.02}). For the admitted Lie
symmetries we determine the algebraic structure and the one-dimensional
optimal system.

\subsection{Lie symmetry analysis for $\Phi _{A}$}

Equation $\Phi _{A}$ admits the Lie symmetry vectors%
\begin{equation}
\Gamma _{1}=\partial _{u}~,~\Gamma _{2}=\partial _{t}~,~\Gamma _{3}=\partial
_{x}~,~\Gamma _{4}=t\partial _{t}+\frac{1}{2}\left( x\partial _{x}-u\partial
_{u}\right) .
\end{equation}%
The commutators of the latter symmetry vectors are presented in \ref{tab1},
while the adjoint representation of the admitted Lie symmetries is given in
Table \ref{tab2}.

\begin{table}[tbp] \centering%
\caption{Commutator table for the admitted Lie point symmetries of equation
$\Phi_A$}%
\begin{tabular}{ccccc}
\hline\hline
$\left[ \Gamma _{i},\Gamma _{j}\right] $ & $\mathbf{\Gamma }_{1}$ & $\mathbf{%
\Gamma }_{2}$ & $\mathbf{\Gamma }_{3}$ & $\mathbf{\Gamma }_{4}$ \\ \hline
$\mathbf{\Gamma }_{1}$ & $0$ & $0$ & $0$ & $-\frac{1}{2}\Gamma _{1}$ \\
$\mathbf{\Gamma }_{2}$ & $0$ & $0$ & $0$ & $\Gamma _{2}$ \\
$\mathbf{\Gamma }_{3}$ & $0$ & $0$ & $0$ & $\frac{1}{2}\Gamma _{3}$ \\
$\mathbf{\Gamma }_{4}$ & $\frac{1}{2}\Gamma _{1}$ & $-\Gamma _{2}$ & $-\frac{%
1}{2}\Gamma _{3}$ & $0$ \\ \hline\hline
\end{tabular}%
\label{tab1}%
\end{table}%

\begin{table}[tbp] \centering%
\caption{Adjoint representation for the admitted Lie point symmetries of
equation $\Phi_A$}%
\begin{tabular}{ccccc}
\hline\hline
$Ad\left( e^{\left( \varepsilon \mathbf{\Gamma }_{i}\right) }\right) \mathbf{%
\Gamma }_{j}$ & $\mathbf{\Gamma }_{1}$ & $\mathbf{\Gamma }_{2}$ & $\mathbf{%
\Gamma }_{3}$ & $\mathbf{\Gamma }_{4}$ \\ \hline
$\mathbf{\Gamma }_{1}$ & $\Gamma _{1}$ & $\Gamma _{2}$ & $\Gamma _{3}$ & $%
\frac{\varepsilon }{2}\Gamma _{1}+\Gamma _{4}$ \\
$\mathbf{\Gamma }_{2}$ & $\Gamma _{1}$ & $\Gamma _{2}$ & $\Gamma _{3}$ & $%
-\varepsilon \Gamma _{2}+\Gamma _{4}$ \\
$\mathbf{\Gamma }_{3}$ & $\Gamma _{1}$ & $\Gamma _{2}$ & $\Gamma _{3}$ & $-%
\frac{\varepsilon }{2}\Gamma _{3}+\Gamma _{4}$ \\
$\mathbf{\Gamma }_{4}$ & $e^{-\frac{\varepsilon }{2}}\Gamma _{1}$ & $%
e^{\varepsilon }\Gamma _{2}$ & $e^{\frac{\varepsilon }{2}}\Gamma _{3}$ & $%
\Gamma _{4}$ \\ \hline\hline
\end{tabular}%
\label{tab2}%
\end{table}%

In order to derive the one-dimensional optimal system we should find the
invariants of the adjoint representation. Indeed the invariants are given by
the system%
\begin{equation}
a_{4}\frac{\partial \Phi }{\partial a_{1}}=0~,~a_{4}\frac{\partial \Phi }{%
\partial a_{2}}=0\text{ and }a_{4}\frac{\partial \Phi }{\partial a_{3}}=0,
\end{equation}%
that is, $\Phi \left( a_{1},a_{2},a_{3},a_{4}\right) =\Phi \left(
a_{4}\right) $ from where it follows that the only invariant is the $a_{4}$.

Hence, let us consider $\mathbf{\Gamma }$ ~to be the generic symmetry vector%
\begin{equation}
\mathbf{\Gamma }=a_{1}\Gamma _{1}+a_{2}\Gamma _{2}+a_{3}\Gamma
_{3}+a_{4}\Gamma _{4},
\end{equation}%
where $a_{1},~a_{2},~a_{3}$ and $a_{4}$ are coefficient constants, then we
should consider the two cases $a_{4}\neq 0$ and $a_{4}=0$.

For $a_{4}\neq 0$ by using Table \ref{tab2} we can write%
\begin{equation}
\mathbf{\Gamma }^{\prime }=Ad\left( e^{\left( \varepsilon _{3}\mathbf{\Gamma
}_{3}\right) }\right) \left( Ad\left( e^{\left( \varepsilon _{2}\mathbf{%
\Gamma }_{2}\right) }\right) \left( Ad\left( e^{\left( \varepsilon _{1}%
\mathbf{\Gamma }_{1}\right) }\right) \Gamma \right) \right)
\end{equation}%
that is%
\begin{equation}
\mathbf{\Gamma }^{\prime }=a_{4}\Gamma _{4}
\end{equation}%
for specific values of $\varepsilon _{1},\varepsilon _{2}$ and $\varepsilon
_{3}$. Hence the similarity transformation which follows from the generic
vector field $\mathbf{\Gamma }$ is equivalent with that of $\Gamma _{4}$.

In the second case where $a_{4}=0$, we have that $\mathbf{\Gamma }%
=a_{1}\Gamma _{1}+a_{2}\Gamma _{2}+a_{3}\Gamma _{3}$, where it can not
simplified more. Hence, we conclude that the one-dimensional optimal system
is consisted by the one-dimensional Lie algebras%
\begin{eqnarray*}
&&\left\{ \Gamma _{1}\right\} ~,~\left\{ \Gamma _{2}\right\} ~,~\left\{
\Gamma _{3}\right\} ~,~\left\{ \Gamma _{4}\right\} ~,~\left\{ \Gamma
_{1}+\gamma \Gamma _{2}\right\} ~,~ \\
&&\left\{ \Gamma _{1}+\gamma \Gamma _{3}\right\} ~,~\left\{ \Gamma
_{2}+\gamma \Gamma _{3}\right\} ~,~\left\{ \Gamma _{1}+\gamma \Gamma
_{2}+\delta \Gamma _{3}\right\} \text{.}
\end{eqnarray*}

\subsubsection{Special case $\protect\alpha =0$}

For $\alpha =0$, the admitted Lie symmetries of equation (\ref{p.01}) are
\begin{equation}
\tilde{\Gamma}_{1}=\partial _{u}~,~\tilde{\Gamma}_{2}=\partial _{t}~,~\tilde{%
\Gamma}_{3}=\partial _{x}~,~\tilde{\Gamma}_{4}=t\partial _{u}~,
\end{equation}%
\begin{equation}
~\tilde{\Gamma}_{5}=t^{2}\partial _{u},~~\tilde{\Gamma}_{6}=-\left(
2t\partial _{t}+x\partial _{x}\right) +u\partial _{u}.
\end{equation}%
Similarly, the commutators and the adjoint representation are given in
Tables \ref{tab3} and \ref{tab4} respectively.

\begin{table}[tbp] \centering%
\caption{Commutator table for the admitted Lie point symmetries of equation
$\Phi_A$ when $\alpha=0$}%
\begin{tabular}{ccccccc}
\hline\hline
$\left[ \Gamma _{i},\Gamma _{j}\right] $ & $\tilde{\Gamma}_{1}$ & $\tilde{%
\Gamma}_{2}$ & $\tilde{\Gamma}_{3}$ & $\tilde{\Gamma}_{4}$ & $\tilde{\Gamma}%
_{5}$ & $\tilde{\Gamma}_{6}$ \\ \hline
$\tilde{\Gamma}_{1}$ & $0$ & $0$ & $0$ & $0$ & $0$ & $\tilde{\Gamma}_{1}$ \\
$\tilde{\Gamma}_{2}$ & $0$ & $0$ & $0$ & $\tilde{\Gamma}_{1}$ & $2\tilde{%
\Gamma}_{4}$ & $-2\tilde{\Gamma}_{2}$ \\
$\tilde{\Gamma}_{3}$ & $0$ & $0$ & $0$ & $0$ & $0$ & $-\tilde{\Gamma}_{3}$
\\
$\tilde{\Gamma}_{4}$ & $0$ & $-\Gamma _{1}$ & $0$ & $0$ & $0$ & $3\tilde{%
\Gamma}_{4}$ \\
$\tilde{\Gamma}_{5}$ & $0$ & $-2\Gamma _{4}$ & $0$ & $0$ & $0$ & $5\tilde{%
\Gamma}_{5}$ \\
$\tilde{\Gamma}_{6}$ & $-\tilde{\Gamma}_{1}$ & $2\tilde{\Gamma}_{2}$ & $%
\tilde{\Gamma}_{3}$ & $-3\tilde{\Gamma}_{4}$ & $5\tilde{\Gamma}_{5}$ & $0$
\\ \hline\hline
\end{tabular}%
\label{tab3}%
\end{table}%

\begin{table}[tbp] \centering%
\caption{Adjoint representation for the admitted Lie point symmetries of
equation $\Phi_A$ when $\alpha=0$}%
\begin{tabular}{ccccccc}
\hline\hline
$Ad\left( e^{\left( \varepsilon \tilde{\Gamma}_{i}\right) }\right) \tilde{%
\Gamma}_{j}$ & $\tilde{\Gamma}_{1}$ & $\tilde{\Gamma}_{2}$ & $\tilde{\Gamma}%
_{3}$ & $\tilde{\Gamma}_{4}$ & $\tilde{\Gamma}_{5}$ & $\tilde{\Gamma}_{6}$
\\ \hline
$\tilde{\Gamma}_{1}$ & $\tilde{\Gamma}_{1}$ & $\tilde{\Gamma}_{2}$ & $\tilde{%
\Gamma}_{3}$ & $\tilde{\Gamma}_{4}$ & $\tilde{\Gamma}_{5}$ & $-\varepsilon
\tilde{\Gamma}_{1}+\tilde{\Gamma}_{6}$ \\
$\tilde{\Gamma}_{2}$ & $\tilde{\Gamma}_{1}$ & $\tilde{\Gamma}_{2}$ & $\tilde{%
\Gamma}_{3}$ & $-\varepsilon \tilde{\Gamma}_{1}+\Gamma _{4}$ & $\varepsilon
^{2}\Gamma _{1}-2\varepsilon \Gamma _{4}+\Gamma _{5}$ & $2\varepsilon \tilde{%
\Gamma}_{2}+\tilde{\Gamma}_{6}$ \\
$\tilde{\Gamma}_{3}$ & $\tilde{\Gamma}_{1}$ & $\tilde{\Gamma}_{2}$ & $\tilde{%
\Gamma}_{3}$ & $\tilde{\Gamma}_{4}$ & $\tilde{\Gamma}_{5}$ & $\varepsilon
\tilde{\Gamma}_{3}+\tilde{\Gamma}_{6}$ \\
$\tilde{\Gamma}_{4}$ & $\tilde{\Gamma}_{1}$ & $\varepsilon \tilde{\Gamma}%
_{1}+\tilde{\Gamma}_{2}$ & $\tilde{\Gamma}_{3}$ & $\tilde{\Gamma}_{4}$ & $%
\tilde{\Gamma}_{5}$ & $-3\varepsilon \tilde{\Gamma}_{4}+\tilde{\Gamma}_{6}$
\\
$\tilde{\Gamma}_{5}$ & $\tilde{\Gamma}_{1}$ & $\tilde{\Gamma}%
_{2}+2\varepsilon \tilde{\Gamma}_{4}$ & $\tilde{\Gamma}_{3}$ & $\tilde{\Gamma%
}_{4}$ & $\tilde{\Gamma}_{5}$ & $-5\varepsilon \tilde{\Gamma}_{5}+\tilde{%
\Gamma}_{6}$ \\
$\tilde{\Gamma}_{6}$ & $e^{\varepsilon }\tilde{\Gamma}_{1}$ & $%
e^{-2\varepsilon }\tilde{\Gamma}_{2}$ & $e^{-\varepsilon }\tilde{\Gamma}_{3}$
& $e^{3\varepsilon }\tilde{\Gamma}_{4}$ & $e^{5\varepsilon }\tilde{\Gamma}%
_{5}$ & $\tilde{\Gamma}_{6}$ \\ \hline\hline
\end{tabular}%
\label{tab4}%
\end{table}%

The invariants of the adjoint representation are determined by the system%
\begin{equation}
a_{6}\frac{\partial \Phi }{\partial a_{1}}=0~,~a_{4}\frac{\partial \Phi }{%
\partial a_{1}}+2a_{5}\frac{\partial \Phi }{\partial a_{4}}-2a_{6}\frac{%
\partial \Phi }{\partial a_{2}}=0~,
\end{equation}%
\begin{equation}
a_{6}\frac{\partial \Phi }{\partial a_{3}}=0~,~-a_{2}\frac{\partial \Phi }{%
\partial a_{1}}+3a_{6}\frac{\partial \Phi }{\partial a_{4}}=0~,~-2a_{2}\frac{%
\partial \Phi }{\partial a_{1}}+5a_{6}\frac{\partial \Phi }{\partial a_{5}}=0
\end{equation}%
\begin{equation}
-a_{1}\frac{\partial \Phi }{\partial a_{1}}+2a_{2}\frac{\partial \Phi }{%
\partial a_{2}}+a_{3}\frac{\partial \Phi }{\partial a_{3}}-3a_{4}\frac{%
\partial \Phi _{4}}{\partial a_{4}}+5a_{5}\frac{\partial \Phi }{\partial
a_{5}}=0
\end{equation}%
from where it follows that for $a_{6}\neq 0$, $\Phi =\Phi \left(
a_{6}\right) $, that is $a_{6}$ is the unique invariant. On the other hand,
when $a_{6}=0,$ the invariants for the adjoint representation of the five
dimensional subalgebra $\left\{ \tilde{\Gamma}_{1},\tilde{\Gamma}_{2},\tilde{%
\Gamma}_{3},\tilde{\Gamma}_{4},\tilde{\Gamma}_{5}\right\} $ are $%
a_{2},a_{3},a_{5}$,

Consequently, the one-dimensional system consists by the Lie algebras%
\begin{eqnarray*}
&&\left\{ \tilde{\Gamma}_{1}\right\} ~,~\left\{ \tilde{\Gamma}_{2}\right\}
~,~\left\{ \tilde{\Gamma}_{3}\right\} ~,~\left\{ \tilde{\Gamma}_{4}\right\}
~,~\left\{ \tilde{\Gamma}_{5}\right\} ,~ \\
&&\left\{ \tilde{\Gamma}_{6}\right\} ~,\left\{ \tilde{\Gamma}_{1}+\gamma
\tilde{\Gamma}_{2}\right\} ~,~\left\{ \tilde{\Gamma}_{1}+\gamma \Gamma
_{3}\right\} ~, \\
&&\left\{ \tilde{\Gamma}_{2}+\gamma \tilde{\Gamma}_{3}\right\} ~,~\left\{
\tilde{\Gamma}_{2}+\gamma \tilde{\Gamma}_{5}\right\} ,~\left\{ \tilde{\Gamma}%
_{3}+\gamma \tilde{\Gamma}_{5}\right\} ~, \\
&&\left\{ \tilde{\Gamma}_{1}+\gamma \tilde{\Gamma}_{2}+\delta \tilde{\Gamma}%
_{3}\right\} ~\ ,~\left\{ \tilde{\Gamma}_{2}+\gamma \tilde{\Gamma}%
_{3}+\delta \tilde{\Gamma}_{4}\right\} ~,~ \\
&&\left\{ \tilde{\Gamma}_{2}+\gamma \tilde{\Gamma}_{3}+\delta \tilde{\Gamma}%
_{5}\right\} .
\end{eqnarray*}

\subsection{Lie symmetry analysis for $\Phi _{B}$}

The application of Lie's algorithm for equation $\Phi _{B}$ provides with
the Lie point symmetries
\begin{equation}
\Gamma _{1}=\partial _{u}~,~\Gamma _{2}=\partial _{t}~,~\Gamma _{3}=\partial
_{x}~,~\check{\Gamma}_{4}=-\left( \alpha +\beta \right) \left( 2t\partial
_{t}+x\partial _{x}\right) +\left( \left( \alpha +\beta \right) u+2x\right)
\partial _{u}.
\end{equation}%
with commutators presented in \ref{tab5}. The adjoint representation of the
admitted Lie algebra is given in Table \ref{tab6}, while when $\alpha +\beta
=0$, the adjoint representation is given in Table \ref{tab7}.

In a similar way with equation $\Phi _{A}$, the one-dimensional optimal
system is consisted by the Lie algebras%
\begin{eqnarray*}
&&\left\{ \Gamma _{1}\right\} ~,~\left\{ \Gamma _{2}\right\} ~,~\left\{
\Gamma _{3}\right\} ~,~\left\{ \check{\Gamma}_{4}\right\} ~,~\left\{ \Gamma
_{1}+\gamma \Gamma _{2}\right\} ~,~ \\
&&\left\{ \Gamma _{1}+\gamma \Gamma _{3}\right\} ~,~\left\{ \Gamma
_{2}+\gamma \Gamma _{3}\right\} ~,~\left\{ \Gamma _{1}+\gamma \Gamma
_{2}+\delta \Gamma _{3}\right\} \text{.}
\end{eqnarray*}

When $\alpha +\beta =0$, the invariants of the adjoint representation are $%
a_{2},~a_{3}$ and $a_{4}$, hence the one-dimensional system is%
\begin{eqnarray*}
&&\left\{ \Gamma _{1}\right\} ~,~\left\{ \Gamma _{2}\right\} ~,~\left\{
\Gamma _{3}\right\} ~,~\left\{ \check{\Gamma}_{4}\right\} ~,~\left\{ \Gamma
_{1}+\gamma \Gamma _{2}\right\} ~,~ \\
&&\left\{ \Gamma _{1}+\gamma \Gamma _{3}\right\} ~,~\left\{ \Gamma
_{2}+\gamma \Gamma _{3}\right\} ~,~\left\{ \Gamma _{1}+\gamma \Gamma
_{2}+\delta \Gamma _{3}\right\} \\
&&\left\{ \Gamma _{2}+\gamma \Gamma _{3}+\delta \check{\Gamma}_{4}\right\}
\text{.}
\end{eqnarray*}

\begin{table}[tbp] \centering%
\caption{Commutator table for the admitted Lie point symmetries of equation
$\Phi_B$}%
\begin{tabular}{ccccc}
\hline\hline
$\left[ \Gamma _{i},\Gamma _{j}\right] $ & $\mathbf{\Gamma }_{1}$ & $\mathbf{%
\Gamma }_{2}$ & $\check{\Gamma}_{3}$ & $\check{\Gamma}_{4}$ \\ \hline
$\mathbf{\Gamma }_{1}$ & $0$ & $0$ & $0$ & $\left( \alpha +\beta \right)
\Gamma _{1}$ \\
$\mathbf{\Gamma }_{2}$ & $0$ & $0$ & $0$ & $-2\left( \alpha +\beta \right)
\Gamma _{2}$ \\
$\mathbf{\Gamma }_{3}$ & $0$ & $0$ & $0$ & $2\Gamma _{1}-\left( \alpha
+\beta \right) \Gamma _{3}$ \\
$\check{\Gamma}_{4}$ & $-\left( \alpha +\beta \right) \Gamma _{1}$ & $%
2\left( \alpha +\beta \right) \Gamma _{2}$ & $-2\Gamma _{1}+\left( \alpha
+\beta \right) \Gamma _{3}$ & $0$ \\ \hline\hline
\end{tabular}%
\label{tab5}%
\end{table}%

\begin{table}[tbp] \centering%
\caption{Adjoint representation for the admitted Lie point symmetries of
equation $\Phi_B$}%
\begin{tabular}{ccccc}
\hline\hline
$Ad\left( e^{\left( \varepsilon \mathbf{\Gamma }_{i}\right) }\right) \mathbf{%
\Gamma }_{j}$ & $\mathbf{\Gamma }_{1}$ & $\mathbf{\Gamma }_{2}$ & $\mathbf{%
\Gamma }_{3}$ & $\check{\Gamma}_{4}$ \\ \hline
$\mathbf{\Gamma }_{1}$ & $\Gamma _{1}$ & $\Gamma _{2}$ & $\Gamma _{3}$ & $%
-\varepsilon \left( \alpha +\beta \right) \Gamma _{1}+\check{\Gamma}_{4}$ \\
$\mathbf{\Gamma }_{2}$ & $\Gamma _{1}$ & $\Gamma _{2}$ & $\Gamma _{3}$ & $%
2\varepsilon \left( \alpha +\beta \right) \Gamma _{2}+\check{\Gamma}_{4}$ \\
$\mathbf{\Gamma }_{3}$ & $\Gamma _{1}$ & $\Gamma _{2}$ & $\Gamma _{3}$ & $%
-2\varepsilon \Gamma _{1}+\varepsilon \left( \alpha +\beta \right) \Gamma
_{3}+\check{\Gamma}_{4}$ \\
$\mathbf{\Gamma }_{4}$ & $e^{\varepsilon \left( \alpha +\beta \right)
}\Gamma _{1}$ & $e^{-2\varepsilon \left( \alpha +\beta \right) }\Gamma _{2}$
& $\frac{e^{\varepsilon \left( \alpha +\beta \right) }-e^{-\varepsilon
\left( \alpha +\beta \right) }}{a+\beta }\Gamma _{1}+e^{-\varepsilon \left(
\alpha +\beta \right) }\Gamma _{3}$ & $\check{\Gamma}_{4}$ \\ \hline\hline
\end{tabular}%
\label{tab6}%
\end{table}%

\begin{table}[tbp] \centering%
\caption{Adjoint representation for the admitted Lie point symmetries of
equation $\Phi_B$ when $\alpha+\beta=0$}%
\begin{tabular}{ccccc}
\hline\hline
$Ad\left( e^{\left( \varepsilon \mathbf{\Gamma }_{i}\right) }\right) \mathbf{%
\Gamma }_{j}$ & $\mathbf{\Gamma }_{1}$ & $\mathbf{\Gamma }_{2}$ & $\mathbf{%
\Gamma }_{3}$ & $\check{\Gamma}_{4}$ \\ \hline
$\mathbf{\Gamma }_{1}$ & $\Gamma _{1}$ & $\Gamma _{2}$ & $\Gamma _{3}$ & $%
\check{\Gamma}_{4}$ \\
$\mathbf{\Gamma }_{2}$ & $\Gamma _{1}$ & $\Gamma _{2}$ & $\Gamma _{3}$ & $%
\check{\Gamma}_{4}$ \\
$\mathbf{\Gamma }_{3}$ & $\Gamma _{1}$ & $\Gamma _{2}$ & $\Gamma _{3}$ & $%
-\varepsilon \Gamma _{1}+\check{\Gamma}_{4}$ \\
$\mathbf{\Gamma }_{4}$ & $\Gamma _{1}$ & $\Gamma _{2}$ & $\varepsilon \Gamma
_{1}+\Gamma _{3}$ & $\check{\Gamma}_{4}$ \\ \hline\hline
\end{tabular}%
\label{tab7}%
\end{table}%

\subsubsection{Special case $\protect\alpha =0$}

For $\alpha =0$, the admitted Lie symmetries of equation (\ref{p.01}) are%
\begin{equation}
\tilde{\Gamma}_{1}=\partial _{u}~,~\tilde{\Gamma}_{2}=\partial _{t}~,~\tilde{%
\Gamma}_{3}=\partial _{x}~,~\tilde{\Gamma}_{4}=t\partial _{u}~,
\end{equation}%
\begin{equation}
~\tilde{\Gamma}_{5}=t^{2}\partial _{u},~~\tilde{\Gamma}_{6}^{\prime
}=-\left( 2t\partial _{t}+x\partial _{x}\right) +\left( \beta u+2x\right)
\partial _{u}.
\end{equation}

The commutator table of the admitted Lie symmetries is presented in \ref%
{tab8}, while the adjoint representation of the admitted Lie symmetries in
given in Table \ref{tab9}. Moreover, the one-dimensional optimal system are
exactly that of equation (\ref{p.01}) for $\alpha =0$, where someone should
replace $\tilde{\Gamma}_{6}$ with $\tilde{\Gamma}_{6}^{\prime }$.

\begin{table}[tbp] \centering%
\caption{Commutator table for the admitted Lie point symmetries of equation
$\Phi_P$ when $\alpha=0$}%
\begin{tabular}{ccccccc}
\hline\hline
$\left[ \Gamma _{i},\Gamma _{j}\right] $ & $\tilde{\Gamma}_{1}$ & $\tilde{%
\Gamma}_{2}$ & $\tilde{\Gamma}_{3}$ & $\tilde{\Gamma}_{4}$ & $\tilde{\Gamma}%
_{5}$ & $\tilde{\Gamma}_{6}^{\prime }$ \\ \hline
$\tilde{\Gamma}_{1}$ & $0$ & $0$ & $0$ & $0$ & $0$ & $\tilde{\Gamma}_{1}$ \\
$\tilde{\Gamma}_{2}$ & $0$ & $0$ & $0$ & $\tilde{\Gamma}_{1}$ & $2\tilde{%
\Gamma}_{4}$ & $-2\tilde{\Gamma}_{2}$ \\
$\tilde{\Gamma}_{3}$ & $0$ & $0$ & $0$ & $0$ & $0$ & $-\tilde{\Gamma}_{3}$
\\
$\tilde{\Gamma}_{4}$ & $0$ & $-\Gamma _{1}$ & $0$ & $0$ & $0$ & $3\tilde{%
\Gamma}_{4}$ \\
$\tilde{\Gamma}_{5}$ & $0$ & $-2\Gamma _{4}$ & $0$ & $0$ & $0$ & $5\tilde{%
\Gamma}_{5}$ \\
$\tilde{\Gamma}_{6}$ & $-\tilde{\Gamma}_{1}$ & $2\tilde{\Gamma}_{2}$ & $%
\tilde{\Gamma}_{3}$ & $-3\tilde{\Gamma}_{4}$ & $5\tilde{\Gamma}_{5}$ & $0$
\\ \hline\hline
\end{tabular}%
\label{tab8}%
\end{table}%
\begin{landscape}
\begin{table}[tbp] \centering%
\caption{Adjoint representation for the admitted Lie point symmetries of
equation $\Phi_B$ when $\alpha=0$}%
\begin{tabular}{ccccccc}
\hline\hline
$Ad\left( e^{\left( \varepsilon \tilde{\Gamma}_{i}\right) }\right) \tilde{%
\Gamma}_{j}$ & $\tilde{\Gamma}_{1}$ & $\tilde{\Gamma}_{2}$ & $\tilde{\Gamma}%
_{3}$ & $\tilde{\Gamma}_{4}$ & $\tilde{\Gamma}_{5}$ & $\tilde{\Gamma}%
_{6}^{\prime }$ \\ \hline
$\tilde{\Gamma}_{1}$ & $\tilde{\Gamma}_{1}$ & $\tilde{\Gamma}_{2}$ & $\tilde{%
\Gamma}_{3}$ & $\tilde{\Gamma}_{4}$ & $\tilde{\Gamma}_{5}$ & $-\varepsilon
\beta \tilde{\Gamma}_{1}+\tilde{\Gamma}_{6}^{\prime }$ \\
$\tilde{\Gamma}_{2}$ & $\tilde{\Gamma}_{1}$ & $\tilde{\Gamma}_{2}$ & $\tilde{%
\Gamma}_{3}$ & $-\varepsilon \tilde{\Gamma}_{1}+\Gamma _{4}$ & $\varepsilon
^{2}\Gamma _{1}-2\varepsilon \Gamma _{4}+\Gamma _{5}$ & $2\varepsilon \beta
\tilde{\Gamma}_{2}+\tilde{\Gamma}_{6}^{\prime }$ \\
$\tilde{\Gamma}_{3}$ & $\tilde{\Gamma}_{1}$ & $\tilde{\Gamma}_{2}$ & $\tilde{%
\Gamma}_{3}$ & $\tilde{\Gamma}_{4}$ & $\tilde{\Gamma}_{5}$ & $%
\,-2\varepsilon \tilde{\Gamma}_{1}+\varepsilon \beta \tilde{\Gamma}_{3}+%
\tilde{\Gamma}_{6}^{\prime }$ \\
$\tilde{\Gamma}_{4}$ & $\tilde{\Gamma}_{1}$ & $\varepsilon \tilde{\Gamma}%
_{1}+\tilde{\Gamma}_{2}$ & $\tilde{\Gamma}_{3}$ & $\tilde{\Gamma}_{4}$ & $%
\tilde{\Gamma}_{5}$ & $-3\varepsilon \beta \tilde{\Gamma}_{4}+\tilde{\Gamma}%
_{6}^{\prime }$ \\
$\tilde{\Gamma}_{5}$ & $\tilde{\Gamma}_{1}$ & $\tilde{\Gamma}%
_{2}+2\varepsilon \tilde{\Gamma}_{4}$ & $\tilde{\Gamma}_{3}$ & $\tilde{\Gamma%
}_{4}$ & $\tilde{\Gamma}_{5}$ & $-5\varepsilon \beta \tilde{\Gamma}_{5}+%
\tilde{\Gamma}_{6}^{\prime }$ \\
$\tilde{\Gamma}_{6}$ & $e^{\varepsilon \beta }\tilde{\Gamma}_{1}$ & $%
e^{-2\varepsilon \beta }\tilde{\Gamma}_{2}$ & $\frac{e^{\varepsilon \beta
}-e^{-\varepsilon \beta }}{\beta }\tilde{\Gamma}_{1}+e^{-\varepsilon \beta }%
\tilde{\Gamma}_{3}$ & $e^{3\beta \varepsilon }\tilde{\Gamma}_{4}$ & $%
e^{5\beta \varepsilon }\tilde{\Gamma}_{5}$ & $\tilde{\Gamma}_{6}$ \\
\hline\hline
\end{tabular}%
\label{tab9}%
\end{table}%
\end{landscape}
\section{Similarity Solutions}

\label{sec4}

We continue our analysis by performing reductions with the use of the\ Lie
symmetries and when it is feasible to determine exact and analytic
similarity solutions.

\subsection{Travel-wave solution for equation $\Phi _{A}$}

Consider the invariants of the Lie symmetry vector $\Gamma _{1}+\gamma
\Gamma _{2}+\delta \Gamma _{3}$, which are $z=x-\frac{\delta }{\gamma }t,$ $%
u=\frac{t}{\gamma }+v\left( x-\frac{\delta }{\gamma }t\right) $. The reduced
equation is a fifth-order ordinary differential equation written as%
\begin{equation}
\delta \gamma ^{2}v^{\left( 5\right) }-\left( \delta ^{3}+\alpha \gamma
^{2}-\left( 2\alpha +\beta \right) \gamma ^{2}v_{z}\right) v_{zzz}+\delta
\left( 2\alpha +\beta \right) \left( v_{zz}\right) ^{2}=0,  \label{p.03}
\end{equation}%
which it can integrated and reduced to a third-order ordinary differential
equation
\begin{equation}
v^{\left( 3\right) }+\left( \alpha +\frac{1}{2}\beta \right) \left(
v_{z}\right) ^{2}-\frac{\delta ^{3}+\alpha \gamma ^{2}}{\delta \gamma ^{2}}%
v_{z}+v_{0}z+v_{1}=0.  \label{p.04}
\end{equation}

\subsubsection{Case $2\protect\alpha +\protect\beta =0$}

When $2\alpha +\beta =0$, equation (\ref{p.04}) can be integrated further
and reduced into the second-order\ ordinary differential equation%
\begin{equation}
v^{\left( 2\right) }-\frac{\delta ^{3}+\alpha \gamma ^{2}}{\delta \gamma ^{2}%
}v+\frac{v_{0}}{2}z^{2}+v_{1}z+v_{2}=0.  \label{p.05}
\end{equation}

Equation (\ref{p.05}) is a linear equation and it is maximally symmetric,
i.e. admits eight Lie point symmetries which means that it can linearized.
The general solution is expressed as follows%
\begin{equation}
v\left( z\right) =c_{1}\exp \left( \frac{z}{\gamma }\sqrt{\delta ^{2}+\alpha
\frac{\gamma ^{2}}{\delta }}\right) +c_{2}\exp \left( -\frac{z}{\gamma }%
\sqrt{\delta ^{2}+\alpha \frac{\gamma ^{2}}{\delta }}\right) +\frac{\left(
v_{0}z^{2}+2v_{1}z+2v_{2}\right) \left( \delta ^{3}+\alpha \gamma
^{2}\right) +2\delta \gamma ^{2}v_{0}}{2\left( \delta ^{3}+\alpha \gamma
^{2}\right) ^{2}}.  \label{p.06}
\end{equation}

\subsubsection{Case $2\protect\alpha +\protect\beta \neq 0$}

In the general scenario for arbitrary parameters $\alpha $ and $\beta $,
equation (\ref{p.04}) admits only one symmetry vector~$\partial _{v}$, while
when $v_{0}=0$ it admits as additional symmetry vector the $\partial _{z}$.

For $v_{0}=0$ and with the use of the autonomous symmetry vector $\partial
_{z}$, equation (\ref{p.04}) can be reduced further to the second-order
ordinary differential equation%
\begin{equation}
y_{vv}+\frac{\left( y_{v}\right) ^{2}}{y}-\left( \left( \frac{\delta }{%
\gamma }\right) ^{2}+\frac{\alpha }{\delta }\right) y^{-1}+\frac{v_{1}}{%
\delta \gamma ^{2}}y^{-2}+\left( a+\frac{\beta }{2}\right) =0,  \label{p.07}
\end{equation}%
where $y=v_{z}$. Equation (\ref{p.07}) can be solved by quadratures, that is%
\begin{equation}
\frac{\sqrt{3}\gamma \sqrt{\delta }ydy}{\sqrt{3\gamma \delta
y_{0}-6v_{1}y+3\left( \alpha \gamma ^{2}+\delta ^{3}\right) y^{2}-\left(
2\alpha +\beta \right) \gamma ^{2}y^{3}}}=v.  \label{p.08a}
\end{equation}

However when $v_{0}\neq 0$ equation (\ref{p.04}) can not be integrated any
more. Therefore, in order to investigate the existence of an analytic
solution we apply the method of singularity analysis. In order to make
equation (\ref{p.04}) autonomous we work with the fourth-order equation%
\begin{equation}
\left( v^{\left( 3\right) }+\left( a+\frac{1}{2}\beta \right) \left(
v_{z}\right) ^{2}-\frac{\delta ^{3}+\alpha \gamma ^{2}}{\delta \gamma ^{2}}%
v_{z}\right) _{z}+v_{0}=0.  \label{p.09}
\end{equation}

The first step of the ARS algorithm for equation (\ref{p.09}) give the
leading-order term $v\left( z\right) =\frac{12}{2\alpha +\beta }\left(
z-z_{0}\right) ^{-1}$. The resonances are given by the polynomial equation%
\begin{equation}
\left( s-1\right) \left( s+1\right) \left( s-4\right) \left( s-6\right) =0,
\label{p.10}
\end{equation}%
from where it follows $s_{1}=1,~s_{2}=-1$, $s_{3}=4$ and $s_{4}=6$.
Consequently, we can write the Laurent expansion%
\begin{equation}
v\left( z\right) =\frac{12}{2\alpha +\beta }\left( z-z_{0}\right) ^{-1}+\chi
_{1}+\chi _{2}\left( z-z_{0}\right) +...+\chi _{I}\left( z-z_{0}\right)
^{-1+I}+...~\text{.}  \label{p.11}
\end{equation}%
We replace in (\ref{p.09}) from where we find that solves the equation for $%
\chi _{1},~\chi _{4}$ and~$\chi _{6}$ to be arbitrary constants, while $\chi
_{2}=\frac{\delta ^{3}+\alpha \gamma ^{2}}{\delta \gamma ^{2}\left( 2\alpha
+\beta \right) }$, $\chi _{3}=0$ etc.

Consider the simplest case where $v_{0}=0$,~$\alpha =\beta =4$ and $\delta
=\gamma =1$. Then, the Laurent expansion (\ref{p.11}) becomes%
\begin{equation}
v\left( z\right) =\left( z-z_{0}\right) ^{-1}+\chi _{1}+\frac{5}{12}\left(
z-z_{0}\right) -\frac{5}{144}\left( z-z_{0}\right) ^{3}+\chi _{6}\left(
z-z_{0}\right) ^{5}-\frac{25}{48384}\left( z-z_{0}\right) ^{7}+...
\end{equation}%
where for $\chi _{1}=\frac{\sqrt{5}}{2}$,$~\chi _{2}=0$, the closed form
solution is $v\left( z\right) =\frac{\sqrt{5}e^{\sqrt{5}z}}{e^{\sqrt{5}z}-1}$%
, which is a kink solution.

We perform the change of variables $v\left( z\right) =\sqrt{\varepsilon }v$,
$\left( v_{0},v_{1}\right) =\sqrt{\varepsilon }\left( v_{0},v_{1}\right) $
in (\ref{p.04}), in order to linearize around$~\varepsilon \rightarrow 0.$
The linearized equation is (\ref{p.05}), from where we determine the
periodic solution (\ref{p.06}) which means that describes the general
solution of (\ref{p.04}) for very small values of $v\left( z\right) $.

\subsection{Scaling solution for equation $\Phi _{A}$}

The application of the Lie symmetry vector $\Gamma _{4}$ provides the
scaling similarity transformation $u=t^{-\frac{1}{2}}v\left( w\right) $
where~$w=xt^{-\frac{1}{2}}$.~The reduced equation can be integrated by one
which provides the fourth-order ordinary differential equation%
\begin{equation}
4wv^{\left( 4\right) }+16v_{www}+4\left( \alpha v-\frac{w^{3}}{3}+\left(
2\alpha +\beta \right) v_{w}\right) v_{ww}+8\left( \alpha +\beta \right)
\left( v_{w}\right) ^{2}-9w^{2}v_{w}-15wv+v_{0}=0.  \label{p.12}
\end{equation}%
which does not admit any Lie point symmetry.

We linearized equation (\ref{p.12}) by applying the transformation $v\left(
w\right) =\sqrt{\varepsilon }v\left( w\right) $,~$v_{0}=\sqrt{\varepsilon }%
v_{0}$; thus it follows%
\begin{equation}
4wv^{\left( 4\right) }+16v_{www}-w^{3}v_{ww}-9w^{2}v_{w}-15wv+v_{0}
\label{p.13}
\end{equation}%
from where we provide the solution%
\begin{eqnarray}
v\left( w\right)  &=&c_{1}\cosh \left( \frac{w^{2}}{4}\right) +c_{2}\sinh
\left( \frac{w^{2}}{4}\right) -\frac{1}{4}e^{-\frac{w^{2}}{4}}\sqrt{\pi }%
c_{3}\left( e^{\frac{w^{2}}{2}}\text{\textrm{erf}}\left( \frac{w}{2}\right) +%
\text{\textrm{erfi}}\left( \frac{w}{2}\right) \right) +  \notag \\
&&+\frac{c_{4}}{12}e^{-\frac{w^{2}}{4}}\left( \frac{4}{w}e^{\frac{w^{2}}{4}}+%
\sqrt{\pi }\left( e^{\frac{w^{2}}{2}}\text{\textrm{erf}}\left( \frac{w}{2}%
\right) -\text{\textrm{erfi}}\left( \frac{w}{2}\right) \right) \right) +
\notag \\
&&-\frac{\sqrt{\pi }v_{0}}{32}e^{-\frac{w^{2}}{4}}\left( e^{\frac{w^{2}}{2}}%
\text{\textrm{erf}}\left( \frac{w}{2}\right) -\text{\textrm{erfi}}\left(
\frac{w}{2}\right) \right) .  \label{p.13a}
\end{eqnarray}

\begin{figure}[tbp]
\centering\includegraphics[width=1\textwidth]{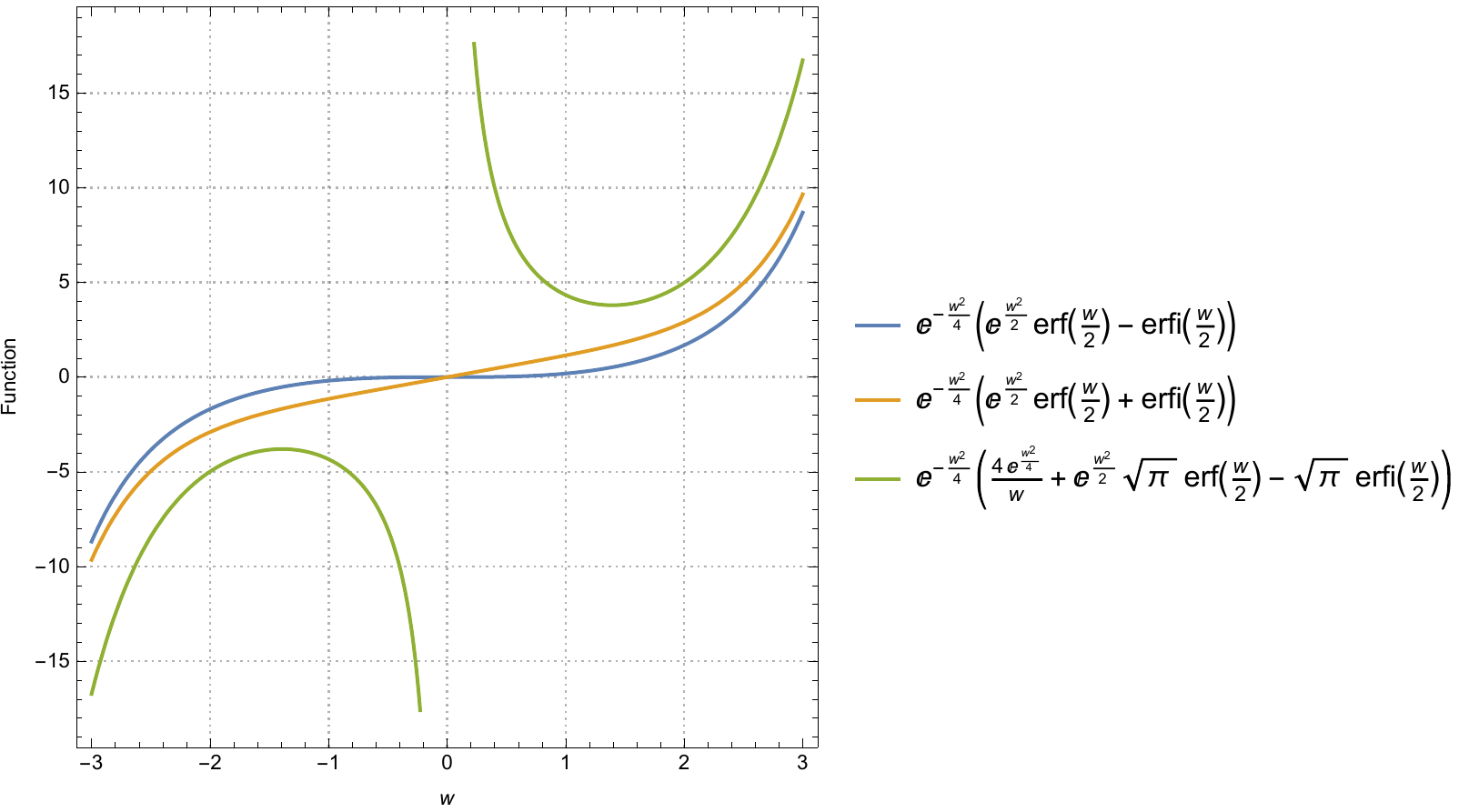}
\caption{The qualitative behaviour of the three non-exponential terms of the
closed-form solution (\protect\ref{p.13a}) are presented. Blue line for the
function $e^{-\frac{w^{2}}{4}}\left( e^{\frac{w^{2}}{2}}\text{\textrm{erf}}%
\left( \frac{w}{2}\right) -\text{\textrm{erfi}}\left( \frac{w}{2}\right)
\right) ,$ orange line is for $e^{-\frac{w^{2}}{4}}\left( e^{\frac{w^{2}}{2}}%
\text{\textrm{erf}}\left( \frac{w}{2}\right) +\text{\textrm{erfi}}\left(
\frac{w}{2}\right) \right) $ and green line is for~the function $e^{-\frac{%
w^{2}}{4}}\left( \frac{4}{w}e^{\frac{w^{2}}{4}}+\protect\sqrt{\protect\pi }%
\left( e^{\frac{w^{2}}{2}}\text{\textrm{erf}}\left( \frac{w}{2}\right) -%
\text{\textrm{erfi}}\left( \frac{w}{2}\right) \right) \right) $}
\label{fig1}
\end{figure}

The qualitative behaviour of the three non-exponential terms of the later
solution are presented in Fig. \ref{fig1}

We apply the same procedure for the determination of similarity solutions
for equation $\Phi _{B}$. We find similar reductions for the travel-wave
solution, and for the scaling solutions. However we study the case where $%
\alpha +\beta =0$ for equation two which provides a new reduction.

\subsection{Travel-wave solution for equation $\Phi _{B}$ and $\protect%
\alpha +\protect\beta =0$}

For equation (\ref{p.02}) with $\alpha +\beta =0$, we apply the Lie
invariants of the symmetry vector $\left\{ \Gamma _{2}+\gamma \Gamma
_{3}+\delta \check{\Gamma}_{4}\right\} $, they are $u\left( t,x\right)
=\delta t\left( x-\frac{\gamma }{2}t\right) +v\left( z\right) $, ~$%
z=x-\gamma t$, while the reduced equation is%
\begin{equation}
\gamma v^{\left( 5\right) }-\left( \gamma \left( \gamma ^{2}-1\right)
+\alpha \delta z-\alpha v_{z}\right) v_{zzz}+\alpha \left( \gamma
v_{zz}-\delta \right) v_{zz}=0,  \label{p.14}
\end{equation}%
Equation (\ref{p.14}) can be integrated twice such that it becomes the
third-order ordinary differential equation%
\begin{equation}
\frac{\delta ^{3}}{\gamma ^{2}}v_{ZZZ}+\frac{\alpha \delta ^{2}}{2\gamma }%
\left( v_{Z}\right) ^{2}-\left( v_{0}+\alpha \delta Z\right) v_{Z}+\alpha
\delta v+v_{0}\frac{v_{0}+\left( 1-\gamma ^{2}+\alpha Z\right) \delta }{%
\alpha \delta ^{2}}+v_{1}=0.  \label{p.15}
\end{equation}%
where we have replaced $z=\frac{v_{0}+\left( 1-\gamma ^{2}+\alpha Z\right)
\delta }{\alpha \delta ^{2}}$.

Equation (\ref{p.15}) admits the point symmetry $\partial _{Z}+\frac{\gamma
}{\delta }t\partial _{w}$. \ With the use of the later lie symmetry,
equation (\ref{p.15}) can be reduced further,
\begin{equation}
\delta ^{5}\alpha Y_{XX}+\alpha ^{2}\delta ^{4}\gamma \sqrt{Y}-2v_{0}\alpha
^{2}\gamma ^{2}+\frac{2\gamma ^{2}}{\sqrt{Y}}\left( \gamma v_{0}\left(
v_{0}+1-\gamma ^{2}\delta \right) +v_{1}\alpha \delta ^{2}+\alpha ^{2}\delta
^{3}X\right) =0.  \label{p.16}
\end{equation}%
where the new variable are $\sqrt{Y}=\frac{\gamma }{\delta }Z-y_{Z}$ and $X=%
\frac{\gamma }{2\delta }W^{2}-y.$ The linearized equation (\ref{p.16}) with
the change of variables $Y\left( X\right) \rightarrow \sqrt{\varepsilon }%
Y\left( X\right) $ is described by the Aire functions.

\section{Conclusion}

\label{sec6}

In this work we studied by using Lie's theory two fifth-order 1+1 partial
differential equations which provide kink solutions. For these two equations
we determine the admitted Lie points symmetries which for the two cases form
a Lie algebra of dimension four, while when the free parameters of the
equations have specific values the admitted Lie algebra has dimension six.

The one-dimensional optimal system is determined for all Lie algebras. We
apply these results to reduce the differential equations and when it is
feasible to write exact and analytic similarity solutions. We found that the
kink solution which was found before \cite{kink1} with the use of the
Hirota's method, can be constructed by using the Lie point symmetries.

The author in \cite{kink1} proposed some families of 2+1 fifth-order partial
differential equations as an extension of his results. The proposed
equations which has kink solutions are%
\begin{equation}
\Phi _{\Gamma }:u_{ttt}-u_{tyyyy}-u_{txx}-\alpha \left( u_{y}u_{ty}\right)
_{y}=0,  \label{p.17}
\end{equation}%
\begin{equation}
\Phi _{\Delta }:u_{ttt}-u_{txxxx}-u_{tyyyy}-\alpha \left( u_{x}u_{tx}\right)
_{x}=0,  \label{p.18}
\end{equation}%
while we introduce the 1+2 fifth-order partial differential equation%
\begin{equation}
\Phi _{E}:u_{ttt}-u_{txxxx}-u_{tyyyy}-u_{txx}-u_{tyy}-\left(
u_{x}u_{tx}\right) _{x}-\left( u_{y}u_{ty}\right) _{y}=0.  \label{p.19}
\end{equation}

We apply the same analysis as before and we find that equation (\ref{p.17})
admits the\ Lie point symmetries
\begin{equation}
\left\{ \Gamma _{1},~\Gamma _{2},~\Gamma _{3},~\Gamma _{4}^{\prime }=\Gamma
_{4}+\frac{1}{2}y\partial _{y},\Gamma _{5}=\partial _{y}\right\}
\end{equation}
and the infinity number of symmetries $\Gamma _{\infty }=U\left( t,x\right)
\partial _{u}$ where $U\left( t,x\right) $ is a solution of the partial
differential equation $U_{ttt}-U_{txx}=0$.

Similarly, equation (\ref{p.17}) is invariant under the one-parameter point
transformations with generators the vector fields $\left\{ \Gamma
_{1},~\Gamma _{2},~\Gamma _{3},~\Gamma _{4}^{\prime },\Gamma _{5}\right\} $
while the infinity number of symmetries is $\Gamma _{\infty }=V\left(
t,y\right) \partial _{u}$ with $V_{ttt}-V_{txx}=0$.

Finally, equation (\ref{p.19}) admits the Lie point symmetries%
\begin{equation}
\left\{ \tilde{\Gamma}_{1},~~\tilde{\Gamma}_{2},~\tilde{\Gamma}_{3},~\tilde{%
\Gamma}_{4},\tilde{\Gamma}_{5},\tilde{\Gamma}_{6}^{^{\prime \prime }}=~%
\tilde{\Gamma}_{6}^{\prime }-y\partial _{y}+2y,\Gamma _{7}=\partial
_{y}\right\} .
\end{equation}

We observe that equations (\ref{p.17}), (\ref{p.18}) and (\ref{p.19}) admit
the Lie symmetries of (\ref{p.01}), (\ref{p.02}) extended in the
two-dimensional flat space $\left\{ x-y\right\} $. Hence, the above results
are also applied and the similarity solutions hold but now in the plane $%
\left\{ x-y\right\} $.

In a future work we plan to investigate the relation of conservation laws
with the existence of kink solutions by using these equations as toy models.

\bigskip%

\end{document}